\documentstyle[psfig]{l-aa}

\begin{document}

\thesaurus{ 05.01.1; 08.08.1; 08.11.1; 08.12.1; 08.16.5 }

\title{ HIPPARCOS results for ROSAT-discovered young stars }

\author{  Ralph Neuh\"auser\inst{1} \and Wolfgang Brandner\inst{2} }

\offprints{Ralph Neuh\"auser}

\institute{Max-Planck-Institut f\"ur extraterrestrische Physik,
D-85740 Garching, Germany; rne@mpe-garching.mpg.de
\and University of Illinois at Urbana-Champaign, Astronomy Dept.,
Urbana IL 61801, USA; brandner@astro.uiuc.edu}

\date {Received 28 August 1997; accepted 5 December 1997}

\maketitle

\markboth{Neuh\"auser \& Brandner:~~~HIPPARCOS results for ROSAT-discovered young stars}{}

\begin{abstract}

Out of $\sim 500$ Lithium-rich ROSAT counterparts, which are presumed 
to be low-mass pre-main sequence stars, 21 stars have been observed 
by HIPPARCOS. We study their parallaxes, proper motions, and photometric data.
For 7 out of 10 Taurus and Lupus stars in our sample, 
proper motions and parallaxes are not inconsistent with 
membership to these associations, while most of the stars 
in Chamaeleon and Scorpius appear to be young foreground stars. 
Combined with ground based photometry and spectroscopy,
HIPPARCOS parallaxes allow us to place 15 stars on an H-R diagram.
All these 15 stars 
are indeed pre-main sequence stars with ages from 1 to 15 Myr. 
Only two of the stars are located on the Hayashi-tracks,
whereas the other 13 are post-T\,Tauri stars located on radiative tracks.
Although this sample is admittedly small,
containing only $3~\%$ of the total sample of Lithium-rich ROSAT 
counterparts, it does not confirm recent predictions by other authors:
We find no stars in the age range from 20 to 100 Myr.
The foreground pre-main sequence stars may have been ejected towards us, 
or they belong to the Gould Belt system, a plane filled with young stars.


\vspace{-4mm}

\end{abstract}

\vspace{-4mm}

\section {Introduction}

Spectroscopic and photometric follow-up observations of 
sources from the ROSAT All-Sky Survey led to the discovery
of many Lithium-rich stars. As Lithium is destroyed gradually in 
the deeper layers of the convection zone (Bodenheimer 1965),
the detection of the Li{\sc I}\,6708\AA\ line in low-resolution
spectra was generally assumed to be a good indicator for the 
youth of a star. About 500 stars were classified as low-mass
pre-main sequence (PMS) stars, and it was assumed that they were located 
at the same distance as previously identified classical T\,Tauri stars 
(cTTS) in the respective star forming regions
(see Neuh\"auser 1997 for a review). As most of these new stars 
lack infrared excess and strong $H \alpha$ emission, they were 
classified as weak-line T\,Tauri stars (wTTS), irregardless of whether 
they are on convective tracks, i.e. being coeval with cTTS, or on
radiative tracks, i.e. being post-TTS. Unlike most cTTS, 
which are associated with dark clouds, the new wTTS show a much more 
wide-spread distribution.

Recently, Brice\~no et al.\ (1997) argued that a dispersed population
of young (foreground) Zero-Age-Main-Sequence (ZAMS) stars with ages 
up to 100 Myr could account for the observed properties of the ROSAT 
sources as well. One can accept a ROSAT counterpart as PMS star only, 
if it shows Li stronger than ZAMS stars of the same spectral type (SpTy).
The Li criterion does not work for G-type stars, 
as G-type 
ZAMS stars still show the primordial Li. 
Membership of a star to an association can be tested by studying
its kinematics, i.e., its radial velocity RV and proper motion PM.
Although RV consistent with membership to an association may be 
suggestive (Neuh\"auser et al.\ 1997, henceforth N97), 
it is not conclusive for showing 3D kinematic membership 
(Frink et al.\ 1997, F97).

A better way to determine the evolutionary status is to measure
the distance to a star, compute its luminosity, and place it on the H-R 
diagram. 
In order to solve the current dispute as to whether the
Li-rich stellar counterparts to ROSAT sources are ZAMS or PMS stars,
we study the HIPPARCOS data (ESA 1997) of 21 Li-rich ROSAT stars,
discuss the kinematics, and present their 
evolutionary status.

\section { Li-rich ROSAT stars observed by HIPPARCOS}
We have compiled a list of all Li-rich ROSAT source counterparts which were
claimed to be PMS stars -- 
found in Chamaeleon (Alcal\'a et al. 1995, A95; Covino et al. 1997, C97), 
Lupus (Krautter et al. 1997, K97; Wichmann et al. 1997b, W97; Wichmann et al.,
in preparation, W98), 
Scorpius (Kunkel et al., in preparation, K98; c.f. also Brandner et al. 1996, B96), 
Taurus (Wichmann et al. 1996, W96; Magazz\`u et al. 1997, M97; N97), 
and Orion (Alcal\'a et al. 1996).
21 of the Lithium-rich ROSAT counterparts have been observed by HIPPARCOS (Table 1).

\section{HIPPARCOS astrometry results}

\begin{table}

\begin{tabular}{llccll}
\hline\noalign{\smallskip}
\multicolumn{6}{c}{\bf Table 1: Lithium-rich ROSAT stars in HIP} \\ \hline
 Designation            & Area&SpTy&Ref.& Li   &Ref. \\ \hline 
 BD$+11 ^{\circ} 533$   & Tau & G2 & M97& 0.10 &M97 (a) \\
 HD 284149              & Tau & G1 & W96& 0.20 &W96 \\
 BD$+17^{\circ} 724$B   & Tau & G5 & W96& 0.41 &W96 \\
 HD 283798              & Tau & G7 & W96& 0.29 &W96 \\ 
 HD 81485               & Cha & G3 & A95& 0.05 &C97 (b) \\
 HD 84075               & Cha & G1 & A95& 0.16 &C97 (b) \\
 HD 92727               & Cha & G1 & A95& 0.03 &C97 (b) \\
 HD 99827               & Cha & F5 & A95& 0.08 &C97 (c) \\
 RXJ1158.5$-$7754a      & Cha & K4 & A95& 0.48 &C97 (c) \\
 RXJ1159.7$-$7601       & Cha & K4 & A95& 0.50 &C97 (c) \\
 RXJ1224.8$-$7503       & Cha & K2 & A95& 0.25 &A95 \\
 HD 109138              & Cha & K1 & A95& 0.13 &C97 (c) \\ 
 HD 137727              & US-B& K0 & B96& 0.13 &K98 \\
 HD 138009              & US-B& G6 & B96& 0.34 &K98 \\
 HD 140637              & US-B& K3 & B96& 0.43 &K98 \\ 
 RXJ1504.8$-$3950       & Lup & F8 & W97& 0.33 &W97 \\
 HD 134974              & Lup & G7 & K97& 0.17 &W98 \\
 HD 141277              & Lup & K0 & K97& 0.60 &K97 \\
 CoD$-36 ^{\circ} 10569$& Lup & K3 & K97& 0.48 &W98 \\
 HD 143677              & Lup & K1 & K97& 0.58 &W98 \\
 HD 143978              & Lup & G2 & K97& 0.17 &W98 \\ \hline
\end{tabular}

Remarks: `US-B' stands for Upper Scorpius B (cf.\ B96).
All stars are classified as `wTTS' (by ref. given for SpTy). 
`Li' is equivalent width $W_{\lambda}$(Li) 
of the Li{\sc I}\,6708\AA\ line in \AA . 
$W_{\lambda}$(Li) for K97 and W96 stars are from R. Wichmann 
(private communication); spectra are shown in K97 and W96. 
SpTy for A95 stars are from C97.
(a) Classified `PMS?' (M97), RV inconsistent with membership (N97).
(b) RV inconsistent with membership (C97).
(c) RV consistent with membership (C97).

\end{table}

To check whether our stars could be members of the respective associations
regarding distances and kinematics, 
we need to know PM, RV, and distances of bona-fide TTS.
HIPPARCOS data of well-known PMS stars are given by Wichmann et al. (1997c).
Averaging the parallaxes of the stars in each region,
they obtain the following distances:
Taurus (5 PMS stars used) at $(142 \pm 14)$\,pc,
Lupus (5) at $(190 \pm 27)$\,pc, and Cha I (3) at $(160 \pm 17)$\,pc.
De Bruijne et al.\ (1997) give $(145 \pm 2)$\,pc as mean
distance to early-type stars in the Upper Sco association.
These values agree with previous distance estimates.
Since all these associations extend by tens of pc in the plane 
of the sky, we have to expect a similar extent in distance.

\begin{figure*}[htb]
\vspace{-0.65cm}
\centerline{\psfig{figure=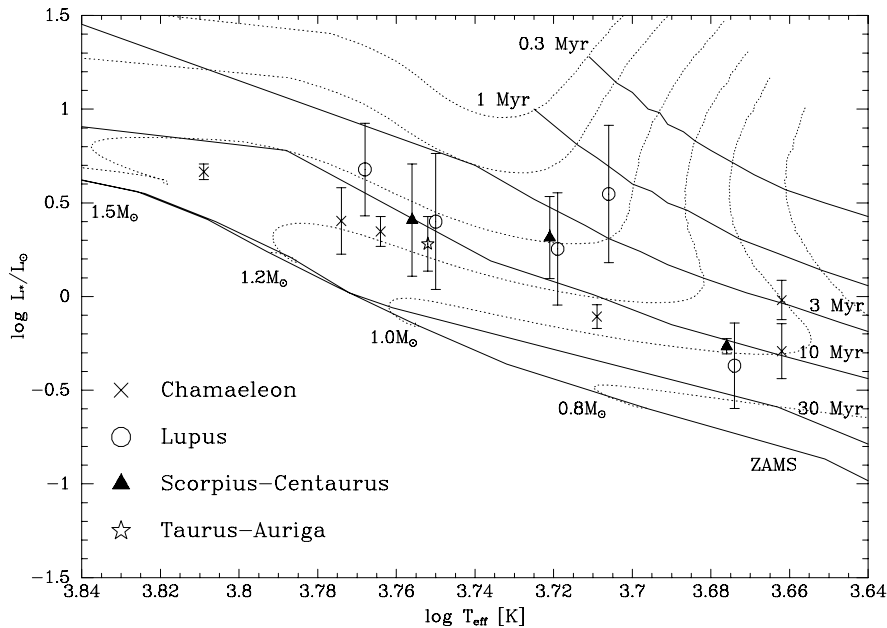,angle=90,width=12.3cm}}
\caption{H-R diagram for our stars with theoretical tracks and isochrones from D'Antona \& Mazzitelli
(1994) with Alexander opacities and CM convection. Error bars are calculated from errors in
the HIPPARCOS parallaxes. }
\end{figure*}

\begin{table*}

\begin{tabular}{lcrrrrllrlrl}
\hline\noalign{\smallskip}
\multicolumn{12}{c}{\bf Table 3: HIPPARCOS astrometry results and physical parameters of stars} \\ \hline
Designation&HIP&\multicolumn{2}{c}{Proper motion [mas/yr]}& 
  Parallax & \multicolumn{2}{c}{Dist.} & $\log$T$_{\rm eff}$ & \multicolumn{2}{c}{$\log$ L/L$_{\odot}$} & Age & Mass \\ 
& no. & $\mu _{\alpha} \cdot \cos \delta$ 
    & $\mu _{\delta}$ & $\pi$~~[mas] & \multicolumn{2}{c}{[pc]}&[K] &  & & [Myr] & $[M_{\odot}]$ \\ \hline 
\noalign{\smallskip}

BD$+11^{\circ}533$ &18117 &$   6.4 \pm  1.8$&$-16.6 \pm  1.5$&$ 6.6 \pm  1.6$&153&\hspace{-3mm}$^{+ 50}_{-30}$& 3.768\\
HD 284149          &19176 &$   6.0 \pm  1.6$&$-15.4 \pm  1.2$&$ 6.4 \pm  1.8$&156&\hspace{-3mm}$^{+ 62}_{-35}$& 3.774\\
BD$+17^{\circ}724$B&20782 &$  -6   \pm 26  $&$-33   \pm 21  $&$ 8   \pm 17  $& (1), & \hspace{-5mm} (2)        & 3.760 \\
HD 283798          &21852 &$  -0.7 \pm  1.4$&$-20.6 \pm  1.0$&$ 8.7 \pm  1.4$&115&\hspace{-3mm}$^{+ 21}_{-16}$& 3.751  & $ 0.28$&\hspace{-5mm} $ ^{+0.15 }_{ -0.13}$& 12 & 1.3\\ \hline

HD 81485B          &45734 &$-107.4 \pm  1.3$&$ 70.4 \pm  1.1$&$13.8 \pm  1.2$& 73&\hspace{-3mm}$^{+  7}_{- 6}$& 3.764  & $ 0.35$&\hspace{-5mm} $ ^{+0.08 }_{ -0.09}$& 14 & 1.2\\
HD 84075           &47135 &$ -72.9 \pm  0.7$&$ 49.8 \pm  0.6$&$15.9 \pm  0.7$& 63&\hspace{-3mm}$^{+  3}_{- 3}$& 3.774\\
HD 92727           &52172 &$   3.3 \pm  1.6$&$ 16.8 \pm  1.4$&$ 7.8 \pm  1.5$&128&\hspace{-3mm}$^{+ 29}_{-20}$& 3.774  & $ 0.40$&\hspace{-5mm} $ ^{+0.18 }_{ -0.15}$& 16 & 1.2\\
HD 99827$^{(3)}$   &55746 &$ -48.0 \pm  0.6$&$ 12.1 \pm  0.6$&$12.1 \pm  0.6$& 83&\hspace{-3mm}$^{+  4}_{- 4}$& 3.809  & $ 0.67$&\hspace{-5mm} $ ^{+0.04 }_{ -0.04}$& 15 & 1.4\\
RXJ1158.5$-$7754a$^{(4)}$&58400 &$ -41.4 \pm  1.4$&$ -0.8 \pm  1.1$&$11.6 \pm  1.3$& 86&\hspace{-3mm}$^{+ 11}_{- 9}$& 3.662  & $-0.02$&\hspace{-5mm} $ ^{+0.11 }_{ -0.10}$& 3 & 1.0\\
RXJ1159.7$-$7601   &58490 &$ -39.9 \pm  1.7$&$ -4.7 \pm  1.5$&$10.8 \pm  1.7$& 92&\hspace{-3mm}$^{+ 17}_{-13}$ & 3.662  & $-0.29$&\hspace{-5mm} $ ^{+0.15 }_{ -0.13}$ & 10 & 1.0\\
RXJ1224.8$-$7503   &60553 &$-240   \pm 20  $&$ -3   \pm 14  $&$41   \pm 18  $& (1)& & 3.690  &     &  \\
HD 109138$^{(3)}$  &61284 &$ -93.3 \pm  1.1$&$ 13.5 \pm  1.1$&$15.2 \pm  1.1$& 66&\hspace{-3mm}$^{+  5}_{- 4}$& 3.706  & $-0.11$&\hspace{-5mm} $ ^{+0.06 }_{ -0.05}$&14&1.1 \\ \hline

HD 137727          &75769 &$  23.9 \pm  2.8$&$ 40.1 \pm  3.1$&$11.5 \pm  2.6$& 87&\hspace{-3mm}$^{+ 25}_{-16}$& 3.720  & $ 0.32$&\hspace{-5mm} $ ^{+0.22 }_{ -0.18}$ & 5 & 1.5\\
HD 138009          &75924 &$ -28.3 \pm  2.8$&$-31.7 \pm  2.8$&$10.9 \pm  3.2$& 92&\hspace{-3mm}$^{+ 38}_{-21}$& 3.756  & $ 0.41$&\hspace{-5mm} $ ^{+0.30 }_{ -0.23}$ & 10 & 1.4\\
HD 140637          &77199 &$ -68.2 \pm  1.1$&$-99.7 \pm  0.9$&$24.4 \pm  1.4$& 41&\hspace{-3mm}$^{+  2}_{- 2}$& 3.675  & $-0.27$&\hspace{-5mm} $ ^{+0.04 }_{ -0.04}$& 11 & 1.0\\ \hline

RXJ1504.8$-$3950   &73777 &$ -30.4 \pm  1.8$&$-32.8 \pm  1.5$&$10.6 \pm  1.7$& 94&\hspace{-3mm}$^{+ 18}_{-13}$& 3.792\\
HD 134974          &74565 &$ -19.8 \pm  2.2$&$-31.4 \pm  1.6$&$ 5.1 \pm  1.8$&195&\hspace{-3mm}$^{+101}_{-50}$& 3.751  & $ 0.40$&\hspace{-5mm} $ ^{+0.36 }_{ -0.26}$& 9 & 1.5\\
HD 141277          &77524 &$ -24.5 \pm  2.0$&$-25.2 \pm  1.8$&$ 6.6 \pm  1.9$&151&\hspace{-3mm}$^{+ 62}_{-34}$& 3.720  & $ 0.25$&\hspace{-5mm} $ ^{+0.30 }_{ -0.22}$& 5 & 1.5\\
CoD$-36^{\circ}10569$&78345&$-16.2 \pm  2.8$&$-48.9 \pm  2.6$&$13.8 \pm  3.2$& 73&\hspace{-3mm}$^{+ 22}_{-14}$&3.675  & $-0.37$&\hspace{-5mm} $ ^{+0.23 }_{ -0.19}$& 15 & 0.95\\
HD 143677          &78684 &$ -14.1 \pm  2.4$&$-25.5 \pm  1.9$&$ 7.0 \pm  2.4$&143&\hspace{-3mm}$^{+ 75}_{-37}$& 3.706  & $ 0.55$&\hspace{-5mm} $ ^{+0.37 }_{ -0.26}$& 1.5 & 1.7\\
HD 143978          &78774 &$ -26.5 \pm  1.4$&$-46.6 \pm  1.3$&$ 6.2 \pm  1.5$&161&\hspace{-3mm}$^{+ 53}_{-32}$& 3.768  & $ 0.68$&\hspace{-5mm} $ ^{+0.25 }_{ -0.19}$& 8 & 1.5\\ \hline
\end{tabular}

Remarks: 
(1) S/N lower than 2.5, i.e., the error in the parallax is too large to give a meaningful distance.
(2) Star B is a few arc seconds off the A-type star BD$+17 ^{\circ}724$, which has
$\pi = (8.0 \pm  1.5)$mas, i.e. $124 ^{+ 29} _{-20}$\,pc.
(3) No V or V-I measurements could be found in the literature.
Thus we rescaled the luminosity estimates given by Alcal\'a et al. (1997).
(4) The companion to this star has SpTy M3 with 
$W_{\lambda}$(Li)$=0.60$ \AA~(C97)
located $\approx 15''$ south of companion {\em a} (A95), i.e.,
most certainly not bound.
\end{table*}

TTS in {\bf Taurus} show a mean PM of 
$(\mu _{\alpha} \cdot \cos \delta , \mu _{\delta}) = (6.4, -22.0)$ mas/yr
(Jones \& Herbig 1979), or $(4.0, -18.7)$ mas/yr, if one includes
the W96 and M97 stars, with $\sim 6.5$ mas/yr scatter (F97). 
While the S/N of the HIPPARCOS data of BD$+17 ^{\circ} 724$B is too low
for any conclusion, the data for the other three Taurus stars are not 
inconsistent with membership and agree well with data given in F97.
Members of {\bf Upper Scorpius} show a mean PM of 
$(\mu _{\alpha} \cdot \cos \delta , \mu _{\delta}) = (-25, -10)$ mas/yr
(de Bruijne et al.\ 1997). 
Space motions and distances indicate that HD 137727, HD 138009, 
and HD 140637 are not kinematic members of 
Upper Sco. Nevertheless, they are young post-TTS with ages
between 5 and 11 Myr (see Table 3).
Eight cTTS in {\bf Lupus} are listed in the STARNET/PPM and show a mean PM of 
$(\mu _{\alpha} \cdot \cos \delta , \mu _{\delta}) = (-12, -26)$ mas/yr
with a scatter of $\sim 7$ mas/yr (Frink et al., in preparation).
Of our six Lupus stars, four have distances consistent with the association, 
and three of those four also have PM consistent with membership, while RXJ1504.8-3950 
and CoD$-36 ^{\circ} 10569$ are not members of the Lupus T association.
All seven stars in {\bf Chamaeleon}, for which we could derive HIPPARCOS 
distances, appear to lie in the foreground of the molecular clouds.
Five stars are young post-TTS with ages between 10 and 16 Myr, 
and RXJ1158.5-7754a is a wTTS with an age of 3 Myr.
HD 89499 (HIP 49616, RXJ10077-8504), classified as a possible PMS star
by A95 and C97, was also observed by HIPPARCOS. Its large PM indicate that
it is a halo star, albeit with an unusual high Li-strength 
(Balachandran et al.\ 1993).
Its parallax is $8.93 \pm 0.73$ mas, i.e. $112 ^{+10} _{-8}$\,pc.

\begin{table}[htb]
\vspace{-0.30cm}
\begin{tabular}{llrrlcc}
\hline\noalign{\smallskip}
\multicolumn{4}{c}{\bf Table 2: Binary stars in the sample } \\ \hline
Designation & $\rho$~~[arc sec]       & $\theta$        & $\Delta H_{p}$~~[mag]
\\ \hline
 HD 81485         & $9.025 \pm 0.005$ & $194 ^{\circ}$  & $1.15 \pm 0.03$ \\
 HD 92727         & $4.981 \pm 0.010$ & $347 ^{\circ}$  & $1.60 \pm 0.05$ \\
 HD 137727        & $2.214 \pm 0.003$ & $183 ^{\circ}$  & $0.31 \pm 0.02$ \\
 HD 138009        & $1.531 \pm 0.006$ & $26  ^{\circ}$  & $0.18 \pm 0.04$ \\
 HD 143677        & $0.290 \pm 0.007$ & $152 ^{\circ}$  & $0.34 \pm 0.08$ \\ \hline
\end{tabular}
\end{table}

\vspace{-5mm}

In Table 2, we list binary companions found by HIPPARCOS: 
Angular separation $\rho$, position angle $\theta$,
and magnitude difference $\Delta H_{p}$ in the HIPPARCOS system.
The HIPPARCOS binary parameters for HD 92727, HD 137727, and HD 138009
agree well with the B96 results from sub-arc second seeing observations 
with SUSI at the ESO-NTT. From the same data set, we also confirm
the parameters measured by HIPPARCOS for the pair HD 81485, although this
wide pair was not included in B96. The companion to HD 140637 could 
not be confirmed, neither by HIPPARCOS nor by near-infrared 
speckle (R. K\"ohler, private communication). Re-examination of 
the SUSI/NTT data shows that the apparent binary companion was an artefact 
due to telescope movement during the exposure. HIPPARCOS also observed the 
A-type star HD 135619 (HIP 74797) in Scorpius, which is located just $18''$ 
off the Li-rich ROSAT star CoD$-34 ^{\circ} 10292$B (K98), a close binary (B96);
the parallax of HD 135619 yields $\approx 124$\,pc,
consistent with Scorpius.
HIPPARCOS lists CoD$-34 ^{\circ} 10292$B as companion to HD 135619.

\section { The H-R diagram based on parallaxes }

Combining SpTy, V, and I (A95, Wichmann 1994, K98)
with HIPPARCOS parallaxes yield the absolute bolometric luminosities of
our stars, assuming intrinsic colours, bolometric corrections, and a
SpTy--T$_{\rm eff}$ relation of late-type dwarfs (Hartigan et al.\ 1994).
We estimate ages and masses by comparison with theoretical
isochrones and tracks from D'Antona \& Mazzitelli (1994).
In Table 3 we present luminosities, ages, and masses.

According to the H-R diagram, shown in Fig. 1, all of our
stars are PMS stars: HD 143677 and RXJ1158.5$-$7754a
are as young as 1.5 Myr to 3 Myr and contract along the Hayashi-tracks.
The other stars have ages between 5 Myr to 15 Myr, 
and are thus post-TTS on radiative tracks.
We find it remarkable that these stars form such a coherent sample
with very similar physical properties (i.e., age and mass) even
though they are spread out over a large region of the sky.
The HIPPARCOS observations confirm earlier PMS classifications 
based mainly on Li and kinematics.
Micela et al.\ (1997) argued that one of their two high-Li EMSS stars 
above the MS according to HIPPARCOS parallaxes is a post-MS giant.
Eleven of our 15 PMS stars, however, do have Li stronger than in giants.
Of the remaining four other stars, at least two have high rotational 
velocities ($v \cdot \sin i \approx 40~km/s$, C97). 
One of them has SpTy later than G0, and thus rotates significantly
faster than a typical post-MS giant (Gray 1989). 

Brice\~no et al. (1997) argued that the majority of the ROSAT sources --
claimed to be PMS stars -- is a dispersed population of young stars 
with ages up to $10^{8}$ yr. Their model assumes continuous star formation
over $10^{8}$ yr, but does not include recent star formation in clouds;
they admit that many Li-rich ROSAT counterparts found on
or near clouds actually are PMS stars. They predict that
there should be roughly three times more $3$ to $10 \times 10^{7}$ yr old 
stars than $\le 3 \cdot 10^{7}$ yr old stars off the clouds, which we cannot 
confirm. All the 15 stars, which we can place on the H-R diagram, 
are $\le 1.6 \cdot 10^{7}$ yr old.

Comparing the $W_{\lambda}$(Li) distribution of these 15 stars 
with the other Li-rich ROSAT sources (Figure 2)
shows that our HIPPARCOS sample -- with the exception of maybe two stars --
is not biased towards high-Li stars.
As only a few of our stars show more Li than Pleiades of the same SpTy,
but all are PMS stars, the Li criterion -- originally used to classify 
these stars as PMS stars -- is conservative. 
Hence, there can be both PMS and ZAMS stars among stars with Li as low 
as ZAMS stars like the Pleiades of the same SpTy, while all stars with 
more Li than ZAMS stars of the same SpTy are younger and therefore PMS stars. 
A few of our $\approx$ 10 Myr old PMS stars show Li even lower than the 
Pleiades (Figure 2), which may indicate that Li depletion is not well 
understood.
However, we cannot completely rule out the possibility that these stars
are unresolved binaries, and thus are erroneously placed too high in the
H-R diagram (cf., e.g., Brandner \& Zinnecker 1997).

\begin{figure}[htb]
\vspace{-0.65cm}
\centerline{\psfig{figure=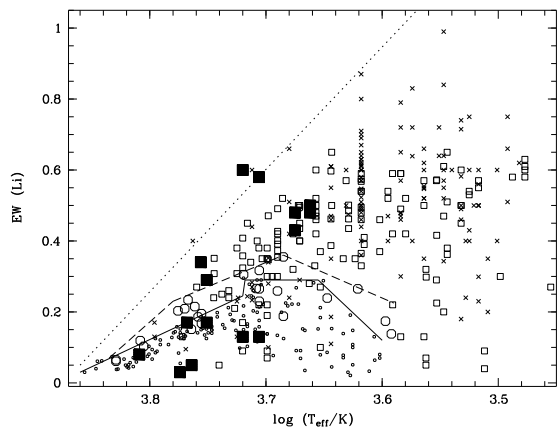,angle=270,width=10.1cm}}
\caption{Lithium equivalent width versus effective temperature:
The stars in our sample with age estimates (filled squares)
are compared to TTS in the Taurus clouds (crosses), 
IC 2602 (large circles), Pleiades (small circles), and Li-rich ROSAT 
counterparts (open squares), c.f.\ N97 for references; 
almost all Li data are from high-resolution spectra. 
The Li content of our PMS stars ranges from the primordial value 
to almost complete depletion.
The lines indicate the upper envelope to Pleiades (full), 
IC 2602 (broken), and T~Tauri stars (dotted).}
\end{figure}

\vspace{-5mm}

Most of the Scorpius and Chamaeleon stars are located foreground 
to the associations, as conjectured by Brice\~no et al. (1997). 
Yet, all of the stars are PMS stars with ages up to 16 Myr.
Unlike cTTS, which are closely associated with dark clouds, 
most of the stars in our sample do not lie in the neighbourhood of a
molecular cloud, but are isolated.
Of the stars with estimated ages, all but two are several degrees
off the nearest clouds; HD 141277 and HD 283798 are projected 
onto the Lup III and L1537 clouds, respectively. 
To investigate whether these isolated PMS stars may have been ejected towards 
us, we need to know their RV. For the Scorpius stars, RV are unknown.
Four of the Chamaeleon stars have RV (C97) consistent with membership,
while the RV of HD 81485B, HD 84075, and HD 92727 indicate that they may 
well have been ejected towards us, i.e. being run-away TTS (Sterzik \& Durisen 1995).
However, given the large errors in the proper motions, it is difficult to 
trace back the place of origin of these $\sim 10$ Myr old post-TTS.
Guillout et al.\ (1998) studied the photometric distances of stars observed by
both ROSAT and TYCHO. By comparing the distance distribution of stars 
on the Belt with those off the Belt, they found evidence for the Gould Belt 
system being a filled plane. Thus, the foreground stars may belong to the 
Gould Belt, if this system is an expanding plane 
-- forming new stars only at its outer edge.

\noindent
{\small\rm
{\em Acknowledgements.\ }
We wish to thank R. Wichmann, M. Kunkel, R. K\"ohler, S. Frink,
J. Alcal\'a, M. Sterzik, and R. Durisen for useful discussion and/or providing 
data prior to publication. RN is supported by the Star Formation program of 
the DFG. WB acknowledges support by Y.-H.\ Chu.}

\end{document}